\documentclass[aps,prd,preprint,preprintnumbers,unsortedaddress,superscriptaddress,showpacs,nofootinbib]{revtex4-1}

\pdfoutput=1

\usepackage{graphicx}
\usepackage{amsmath}
\usepackage{amsfonts}
\usepackage{amssymb}
\usepackage{color}

\usepackage[colorlinks, citecolor=blue,anchorcolor=red,menucolor=red, linkcolor=red,filecolor=red,runcolor=red,urlcolor=blue,frenchlinks=red]{hyperref}

\begin{document}
\arraycolsep1.5pt

\title{Disclosing $D^*\bar{D}^*$ molecular states in the $B_c^- \to \pi^- J/\psi \omega$ decay}

\author{L.~R.~Dai}
\email{dailr@lnnu.edu.cn}
\affiliation{Department of Physics, Liaoning Normal University, Dalian 116029, China}
\affiliation{Departamento de F\'isica Te\'orica and IFIC, Centro Mixto Universidad de Valencia-CSIC,
Institutos de Investigac\'ion de Paterna, Aptdo. 22085, 46071 Valencia, Spain
}

\author{J.~M.~Dias}
\email{jdias@if.usp.br}
\affiliation{Departamento de F\'isica Te\'orica and IFIC, Centro Mixto Universidad de Valencia-CSIC,
Institutos de Investigac\'ion de Paterna, Aptdo. 22085, 46071 Valencia, Spain
}
\affiliation{Instituto de F\'{i}sica, Universidade de S\~{a}o Paulo, Rua do Mat\~{a}o, 1371, Butant\~{a}, CEP 05508-090, S\~{a}o Paulo, S\~{a}o Paulo, Brazil}

\author{E.~Oset}
\email{Eulogio.Oset@ific.uv.es}
\affiliation{Departamento de F\'isica Te\'orica and IFIC, Centro Mixto Universidad de Valencia-CSIC,
Institutos de Investigac\'ion de Paterna, Aptdo. 22085, 46071 Valencia, Spain
}

\date{\today}
\begin{abstract}
We study the $B_c^- \to \pi^- J/\psi \omega$ and $B_c^- \to \pi^- D^* \bar{D}^*$ reactions and show that they
are related by the presence of two resonances, the $X(3940)$ and $X(3930)$, that are of molecular nature
and couple most strongly to $D^* \bar{D}^*$, but also to $J/\psi\omega$. Because of that, in the
$J/\psi\omega$ mass distribution we find a cusp with large strength at the $D^* \bar{D}^*$ threshold and
predict the ratio of strengths between the peak of the cusp and the maximum of the $D^* \bar{D}^*$
distribution close to $D^* \bar{D}^*$ threshold, which are distinct features of the molecular nature of
these two resonances.
\end{abstract}

\maketitle

\section{Introduction}
\label{sec:intro}

Molecular states of mesons have long been the subject of study in hadron physics.
Detailed recent reviews can be seen in Refs.~\cite{slzhu,guoren}. As commented in Ref.~\cite{rosner}
the support for hadron molecules is quite obvious once we realize that baryon molecules
exist in the form of nuclei. In fact, multi-mesons states, not just meson-meson molecules,
have also been advocated, like multi-rho states in Ref.~\cite{luismulti}, $K^*$-multi-rho states in
Ref.~\cite{yamagata}, $D^*$-multi-rho states in Ref.~\cite{xiao}, two mesons and a baryon states
\cite{alberto,enyo} and many others (see a recent review in Ref.~\cite{aphyspo}). Actually, the
interaction between mesons, particularly vector mesons in spin two, is very strong \cite{molina1,gengra,gencpc},
even stronger than between nucleons, and the only limit to the formation of multi-meson
states is that we do not have the meson number conservation, unlike baryon number conservation
for the nucleons forming nuclei. This allows the multi-meson states to decay in states of fewer,
or lighter mesons, the width increases with the number of mesons of the cluster, and at some
point they are no longer identifiable experimentally. Even then, according to \cite{luismulti,yamagata,xiao}, states
up to $6$ vector mesons can be detected and the $f_6(2510)$ qualifies as a six-rho meson
state \cite{luismulti}.

The identification of states as being of molecular nature is not an easy task,
and in general standard quark structures, or multiquark states are competing in the interpretation
\cite{slzhu,rosner}. Yet, there are several experimental features that reveal the molecular
structure \cite{guoren} and ultimately it is the systematic and correct description of experimental
features and the accuracy of the predictions what builds up in favor of this structure for many
states.

The weak decay of heavy mesons and baryons has turned out into one important tool to
identify states of molecular type \cite{osetweak}. Curiously, an interaction that does not
respect parity and isospin, has shown itself as a great tool to identify molecular states because
certain decays filter good quantum numbers due to selection rules, like Cabibbo and
color enhancement in some topologies of decay modes.

One of the features attached to the molecular states that couple to several hadron-hadron
channels, is that by looking at one of the channels with relatively small strength one finds a
strong and unexpected cusp in the threshold of the channels corresponding to the main
component of the molecule. One recent example of this was found in the
$B^+ \to J/\psi \phi K^+$ reaction measured at LHCb \cite{lhcb,lhcbis}. The reaction was analyzed
in \cite{lhcb,lhcbis} and at low invariant masses only the $X(4140)$ state was included,
concluding that its width had to be considerably larger than the average of the PDG \cite{pdg}
from other experiments. A different interpretation, with a better fit to the data, was given in
\cite{wang}, where, in addition to the $X(4140)$, the $X(4160)$ was included in the fit,
assuming that this state is the $D_s^*\bar{D}_s^*$ state predicted in \cite{molina} as a
$0^+[2^{++}]$ state. It is worth noting that other works have also suggested a bound state
of $D_s^*\bar{D}_s^*$ \cite{xliu,thomas,chen,karliner}, although it was originally associated
to the $X(4140)$. This bound $D_s^*\bar{D}_s^*$ state also couples to other light
vector states and to $J/\psi\phi$, hence, it can be observed in this latter channel. However,
the fact that the resonance couples most strongly to $D_s^*\bar{D}_s^*$ has the consequence
that the $J/\psi\phi$ mass spectrum develops a strong cusp at the $D_s^*\bar{D}_s^*$ threshold,
something visible in the experimental spectra with an increased strength in that region. It is also worth
mentioning that a similar enhancement is seen, although with poor statistics, in the recent
BESIII work on the $e^+e^- \to \gamma J/\psi \phi$ reaction \cite{bes3}.

In the present work we want to continue along this line of research and present results for a
reaction that should reveal the $D^*\bar{D}^*$ nature of two states found in \cite{molina} as
$0^+[0^{++}]$ and $0^+[2^{++}]$ at $3943$ MeV and $3922$ MeV, respectively, which can be
identified with some experimental states in that region \cite{molina,pdg} \footnote{The state at
$3943$ MeV can be associated with the $X(3940)$ of \cite{belle,belle2} and the $X(3922)$
with the $X(3930) $\cite{belle3} now classified in the PDG as the $\chi_{c2}(2P)$.}. In this case we note that
the states found, mostly $D^*\bar{D}^*$ bound states, also couple to $J/\psi\omega$ in the second place, and
$J/\psi\phi$ with smaller strength. So we choose the $J/\psi\omega$ observation channel
looking into the necessary cusp that should develop at the $D^*\bar{D}^*$ threshold. For this
purpose we look into the $B_c^-\to \pi^-J/\psi \omega$ decay and then into the $J/\psi \omega$
invariant mass distribution. The choice of this reaction is that in a first stage of the reaction the
$D^*\bar{D}^*$ state is formed with a dominant weak decay mechanism, but the $J/\psi \omega$
state is not formed at this level. Then the $J/\psi \omega$ is finally produced via rescattering of the
$D^*\bar{D}^*$ component with the other components that make up the two molecular states. This
stresses the role of the resonance since there is no tree level $J/\psi\omega$ contribution. Thus, we
obtain two peaks in the $J/\psi\omega$ mass distribution corresponding to the molecular states
and a strong cusp at the $D^*\bar{D}^*$ threshold. In addition we also look at the $D^*\bar{D}^*$ mass
distribution in the $B_c^-\to \pi^-D^*\bar{D}^*$ reaction and evaluate its strength above the $D^*\bar{D}^*$
threshold, which is closely connected to the strength of the $J/\psi\omega$ mass distribution. The
$D^*\bar{D}^*$ cusp feature, together with the relative strength of the $D^*\bar{D}^*$ compared to the one
of $J/\psi\omega$, are two magnitudes which are tied to the molecular structure of these two
resonances, and we encourage the performance of the experiment that should bring valuable
light into these issues.

\section{Formalism}
\label{sec:form}

\begin{figure}
\includegraphics[width=0.45\textwidth]{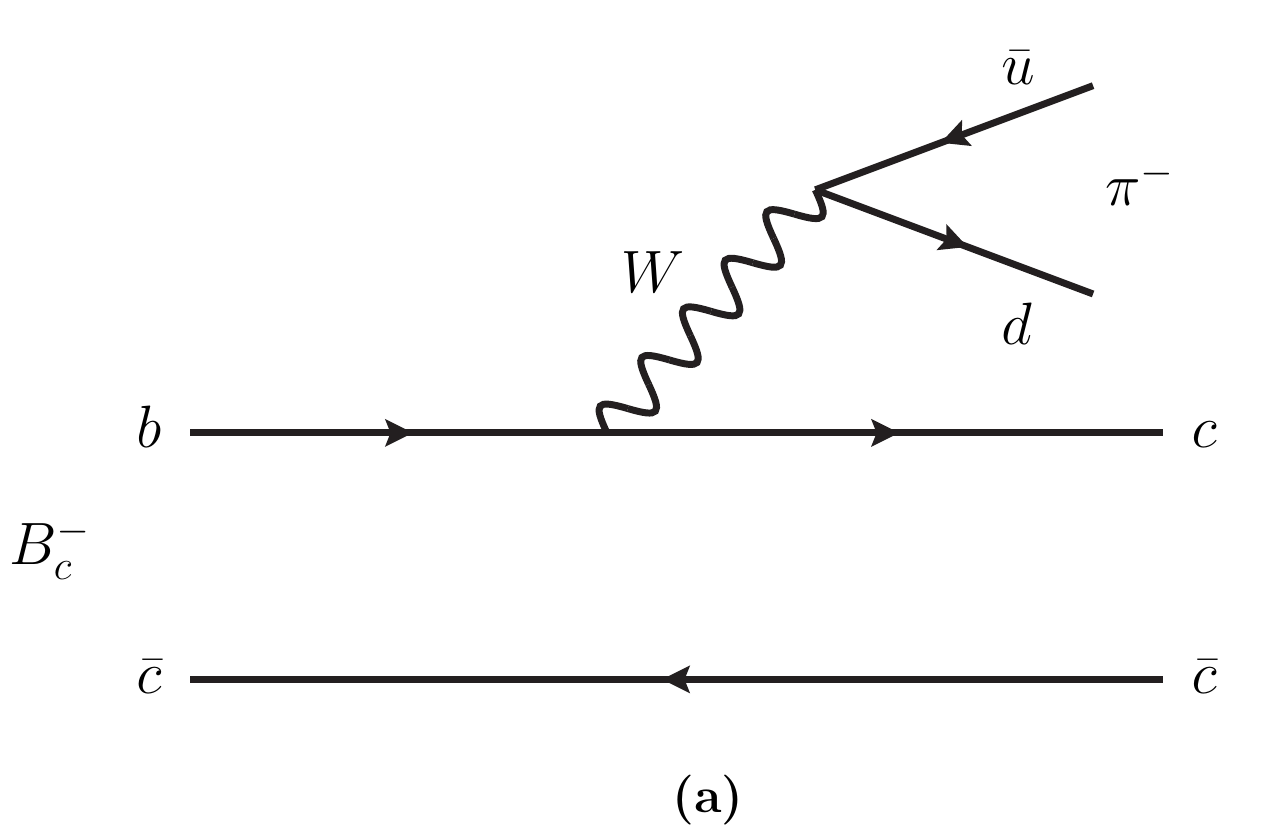} ~~~~	\includegraphics[width=0.45\textwidth]{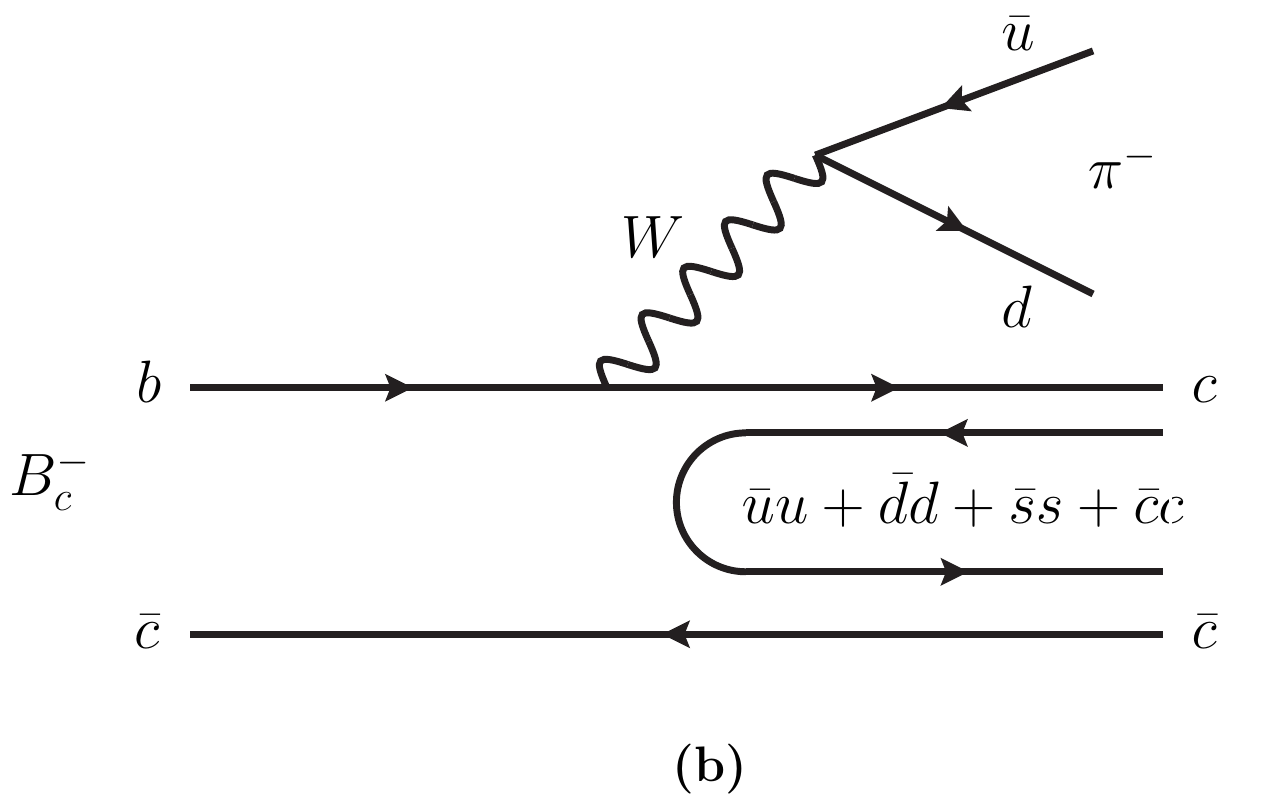}
\caption{\label{Bcdecay} (a) Microscopic quark picture of
	$B_c^- \to \pi^- c\bar{c}$ decay; (b) Hadronization through $\bar{q}q$
	creation with vacuum quantum numbers.}
\end{figure}

We look into the $B_c^-$ decay mechanism at quark level depicted in Fig.~\ref{Bcdecay}(a).
The mechanism qualifies as external emission \cite{chau} and is both Cabibbo favored in
$Wud$ vertex, and color favored (the $Wbc$ vertex is also the least Cabibbo suppressed
of the $b$ decays). Then the final $c$ quark from $b$ decay and the spectator $\bar{c}$
quark from the $B_c^-$ hadronize, with the incorporation of $\bar{q}q$ with the quantum
numbers of the vacuum (see Fig.~\ref{Bcdecay}(b)) to give two mesons. The resulting two mesons are easily
obtained by writing
\begin{equation}\label{had}
H=\sum\limits_{i=1}^{4} c \bar{q}_i q_i \bar{c} =\sum\limits_{i=1}^{4} M_{4i}\,M_{i4}=(M^2)_{44}\nonumber\, ,
\end{equation}
where $M_{ij}$ is the $q\bar{q}$ matrix with the $u,\,d,\,s,\,c$ quarks. However, it is convenient
to write the $q\bar{q}$ matrix in terms of physical mesons, in this case vector mesons as
\begin{equation}
\label{Vmatrix}
M_{ij} \to V =
\left(
\begin{array}{cccc}
\frac{1}{\sqrt{2}} \rho^0 + \frac{1}{\sqrt{2}} \omega & \rho^+ & K^{* +} & \bar{D}^{* 0} \\
\rho^- & -\frac{1}{\sqrt{2}} \rho^0 + \frac{1}{\sqrt{2}} \omega & K^{* 0} & \bar{D}^{* -} \\
K^{* -} & \bar{K}^{* 0} & \phi & D_s^{* -} \\
D^{* 0} & D^{* +} & D_s^{* +} & J/\psi\\
\end{array}
\right)\, ,
\end{equation}
and we get
\begin{equation}
| H \rangle = D^{*0}\bar{D}^{* 0} + D^{*+}\bar{D}^{*-} + D_s^{*+}\bar{D}_s^{*-} + J/\psi J/\psi \, .
\end{equation}
The intrinsic phase convention for isospin multiplets in $(D^{*+},\, -D^{*0})$, $(\bar{D}^{*0},\, D^{*-})$
indicates that the isospin combination of $H$ is $I=0$, as it should be since it comes from $c\bar{c}$.
Thus, we can write
\begin{equation}\label{h}
| H \rangle = \sqrt{2} | D^* \bar{D}^* \rangle + | D_s^* \bar{D}_s^*\rangle \, ,
\end{equation}
where we have neglected the $ J/\psi J/\psi$ component which is far beyond in energy from our range of
concern. In addition, the coupling of the resonances found in \cite{molina} to $J/\psi J/\psi$ is
negligibly small.
\begin{figure}
	\begin{center}
		\includegraphics[width=0.35\textwidth]{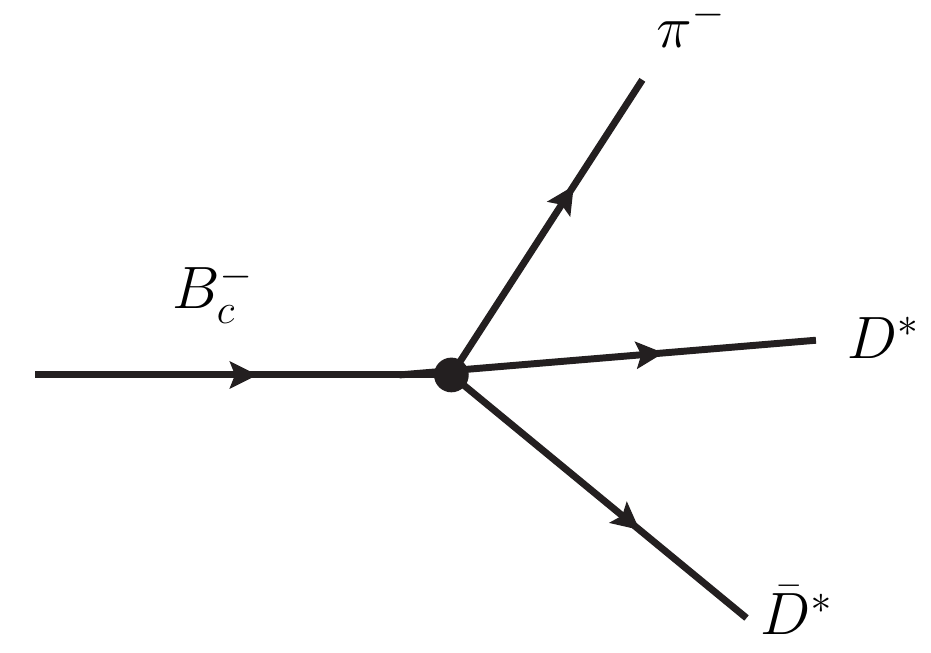}
	\end{center}
	\caption{\label{Bcdecay2} Tree level contribution corresponding to the hadronization depicted in
	Fig.~\ref{Bcdecay}(b).}
\end{figure}

The combination of $| H \rangle$ in Eq.~\eqref{h} accounts only for the flavor composition. We need
to take into account the spin-angular momentum structure of the vertices. If we produce a
$0^+[0^{++}]$ $D^*\bar{D}^*$ state we have $0^- \to 0^-\, 0^+$ transition and we adopt the common choice
of taking the lowest possible angular momentum in the vertex, $L=0$. The s-wave and the
$J=1^-$ of the $D^*$ leads us to a vertex of the type
\begin{equation}\label{A}
A^{\prime}\, \vec{\epsilon} \cdot \vec{\epsilon}^{\,\,\prime}\, ,
\end{equation}
with $\vec{\epsilon}$, $\vec{\epsilon}^{\,\,\prime}$ the polarization vertices of $D^*$, $\bar{D}^*$. Note
that we shall work in the rest frame of the resonances produced, where $D^*$, $\bar{D}^*$ momenta are
small with respect to their masses and then we neglect the $\epsilon^0$ component.
On the other hand, if we produce a $2^{++}$ state, the $0^- \to 0^- \, 2^+$ requires $L=2$ and
we shall then take the D-wave structure
\begin{equation}
B\, (\vec{\epsilon} \cdot \vec{k} \vec{\epsilon}^{\,\,\prime}\cdot \vec{k} -
\frac{1}{3} |\vec{k}|^2\, \vec{\epsilon} \cdot \vec{\epsilon}^{\,\,\prime}) \, ,
\end{equation}
where $\vec{k}$ is the momentum of the pion. Hence, the tree level amplitude for $B_s^- \to \pi^- D^*\bar{D}^*$
shown in Fig. 2 is given by
\begin{equation}\label{tree}
t^{tree}_{B_c \to \pi^- D^* \bar{D}^*} = \sqrt{2}\Big[ A\, |\vec{k}_{av}|^2\, \vec{\epsilon}\cdot \vec{\epsilon}^{\,\,\prime}
+ B(\vec{\epsilon} \cdot \vec{k} \,\, \vec{\epsilon}^{\,\,\prime}\cdot \vec{k} -
 \frac{1}{3} |\vec{k}|^2\, \vec{\epsilon} \cdot \vec{\epsilon}^{\,\,\prime})\Big] \, ,
\end{equation}
where we have substituted $A^{\prime}$ of Eq.~\eqref{A} by $A\, |\vec{k}_{av}|^2$, with
$\vec{k}_{av}$, an average value of $\vec{k}$, just to make $A$ and $B$ have the same dimension.
We take $|\vec{k}_{av}|=1000$ MeV.

After the first step for $D^*\bar{D}^*$ and $D_s^*\bar{D}_s^*$ production, these mesons undergo
final state interaction, as depicted in Fig. 3 and 4, to produce $J/\psi\omega$ and $D^*\bar{D}^*$
in the final state. In the case of $J/\psi\omega$ production shown in Fig. 3, since this state
is not primarily produced in $| H \rangle$, it is produced through rescattering via the resonances
$X(3922)$ and $X(3943)$. In the case of $D^*\bar{D}^*$ production, shown in Fig. 4, it proceeds via
tree level (primary production, Fig. 4(a)) and rescattering (Fig. 4(b)).

\begin{figure}
	\begin{center}
		\includegraphics[width=0.48\textwidth]{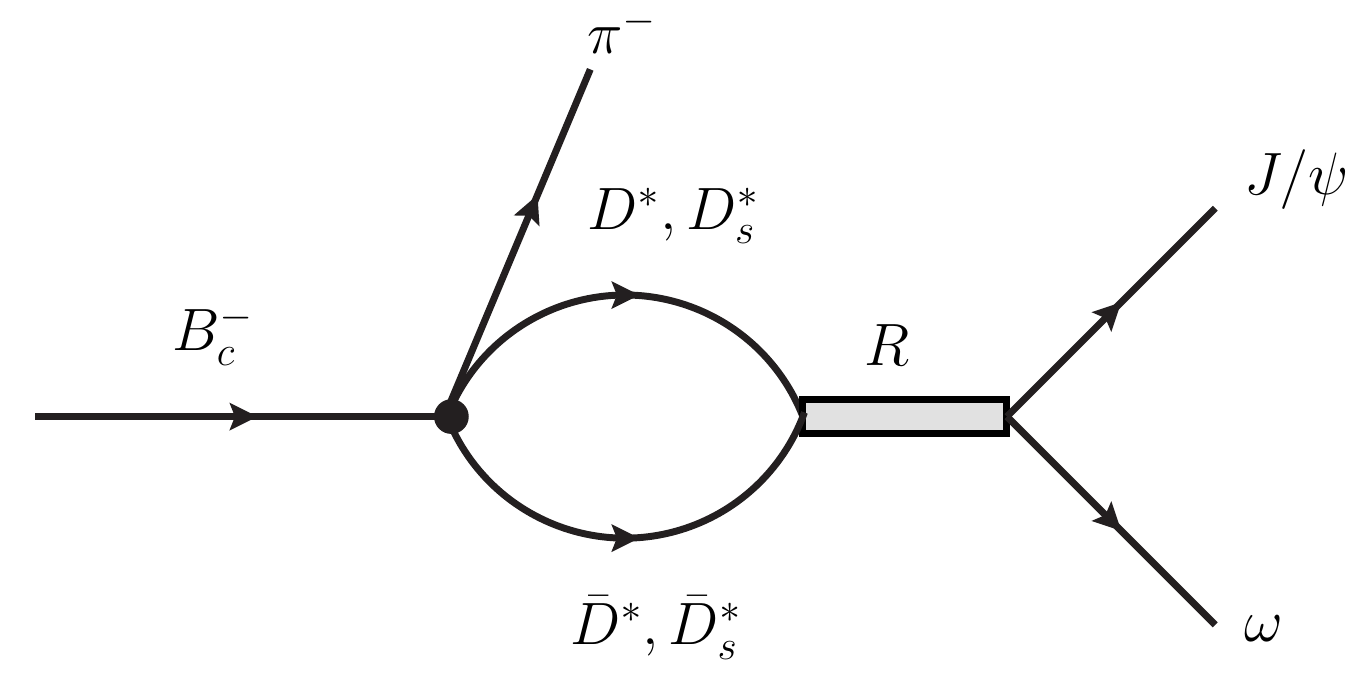}
	\end{center}
	\caption{\label{Bcloop} Mechanism to produce the $J/\psi\omega$ final state
	through rescattering of the $D^*\bar{D}^*$ and $D_s^*\bar{D}_s^*$ components. $R$
	is either the $X(3922)\,(2^{++})$ or $X(3943)\,(0^{++})$.}
\end{figure}

\begin{figure}
	\begin{center}
		\includegraphics[width=0.72\textwidth]{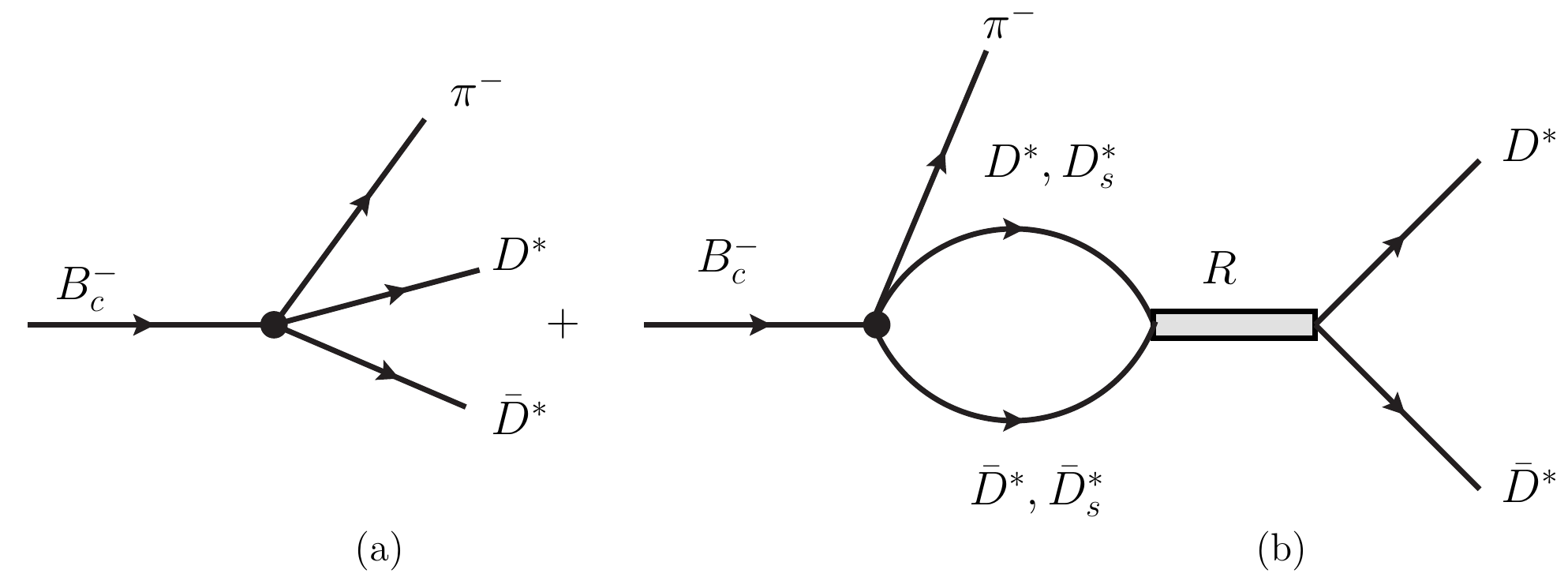}
	\end{center}
	\caption{\label{Bcdiag} Mechanism to produce the $D^*\bar{D}^*$ in the final state
	through tree level (a) and rescattering (b). $R$ is either the $X(3922)\,(2^{++})$
	or $X(3943)\,(0^{++})$.}
\end{figure}

Analytically, we have
\begin{equation}\label{jomega}
t_{J/\psi\omega} = A\, |\vec{k}_{av}|^2\, \vec{\epsilon}\cdot \vec{\epsilon}^{\,\,\prime}\,\, t_1 +
B (\vec{\epsilon} \cdot \vec{k} \,\,\vec{\epsilon}^{\,\,\prime}\cdot \vec{k} -
 \frac{1}{3} |\vec{k}|^2\, \vec{\epsilon} \cdot \vec{\epsilon}^{\,\,\prime})\,\, t_2 \, ,
\end{equation}
where
\begin{equation}\label{t1}
t_1 = \sqrt{2} G_{D^*\bar{D}^*}(M_{inv}^{J/\psi\omega}) \,\,
t^{I}_{D^*\bar{D}^*\to J/\psi\omega}(M_{inv}^{J/\psi\omega}) + G_{D_s^*\bar{D}_s^*}(M_{inv}^{J/\psi\omega})
\,\, t^{I}_{D_s^*\bar{D}_s^*\to J/\psi\omega}(M_{inv}^{J/\psi\omega})\, ,
\end{equation}
and
\begin{equation}\label{t2}
t_2 = \sqrt{2} G_{D^*\bar{D}^*}(M_{inv}^{J/\psi\omega}) \,\,
t^{II}_{D^*\bar{D}^*\to J/\psi\omega}(M_{inv}^{J/\psi\omega}) + G_{D_s^*\bar{D}_s^*}(M_{inv}^{J/\psi\omega})
\,\, t^{II}_{D_s^*\bar{D}_s^*\to J/\psi\omega}(M_{inv}^{J/\psi\omega}) \, ,
\end{equation}
while for $D^*\bar{D}^*$ production we have
\begin{equation}\label{dd}
t_{D^*\bar{D}^*} = A\, |\vec{k}_{av}|^2 \, \vec{\epsilon} \cdot \vec{\epsilon}^{\,\,\prime}\, t_3 +
B (\vec{\epsilon} \cdot \vec{k} \,\,\vec{\epsilon}^{\,\,\prime}\cdot \vec{k} -
 \frac{1}{3} |\vec{k}|^2\, \vec{\epsilon} \cdot \vec{\epsilon}^{\,\,\prime})\,\, t_4 \, ,
\end{equation}
with
\begin{equation}\label{t3}
t_3 = \sqrt{2} + \sqrt{2} G_{D^*\bar{D}^*}(M_{inv}^{D^*\bar{D}^*})\,\,
t^I_{D^*\bar{D}^*\to D^*\bar{D}^*}(M_{inv}^{D^*\bar{D}^*}) +
G_{D_s^*\bar{D}_s^*}(M_{inv}^{D^*\bar{D}^*})\,\,
t^I_{D_s^*\bar{D}_s^*\to D^*\bar{D}^*}(M_{inv}^{D^*\bar{D}^*})\, ,
\end{equation}
and
\begin{equation}\label{t4}
t_4 = \sqrt{2} + \sqrt{2} G_{D^*\bar{D}^*}(M_{inv}^{D^*\bar{D}^*})\,\,
t^{II}_{D^*\bar{D}^*\to D^*\bar{D}^*}(M_{inv}^{D^*\bar{D}^*}) +
G_{D_s^*\bar{D}_s^*}(M_{inv}^{D^*\bar{D}^*})\,\,
t^{II}_{D_s^*\bar{D}_s^*\to D^*\bar{D}^*}(M_{inv}^{D^*\bar{D}^*})\, ,
\end{equation}
where $I$, $II$ stand for the $0^{++}$ and $2^{++}$ states, respectively. Since
the $\vec{\epsilon}\cdot \vec{\epsilon}^{\,\,\prime}$ and $\vec{\epsilon}\cdot \vec{k}\,
\vec{\epsilon}^{\,\,\prime}\cdot \vec{k}-\frac{1}{3}|\vec{k}|^2\vec{\epsilon}\cdot \vec{\epsilon}^{\,\,\prime}$
structures filter spin $0$ and $2$ respectively, the structure is kept in the iterations
implicit in Eqs.~\eqref{t1}, \eqref{t2}, \eqref{t3} and \eqref{t4}. The $G$ functions in the former
equations are the vector vector loop functions for the intermediate $D^*\bar{D}^*$, $D_s^*\bar{D}_s^*$
in Figs. 3 and 4(b). They are regularized in Ref.~\cite{molina} using dimensional regularization
with the subtraction constant $a=-2.07$ and $\mu=1000$ MeV. Here, we follow the prescription
of Refs.~\cite{wu,wang} and we use the cutoff method with ${q}_{max}$ fixed to reproduce the results of Ref.~\cite{molina}.
 In the former equations $A$ and $B$
are functions (we take them as constants in the limited range of invariant mass studied) which have to do
with the weight of the weak process and hadronization before the final state interaction is
taken into account. We shall vary $A$ and $B$ within a reasonable range to see the results.

With the amplitudes of Eqs.~\eqref{jomega} and \eqref{dd} the mass distributions,
summing $|t|^2$ over the final vector polarizations, given by
\begin{equation}
\frac{d\Gamma}{dM_{inv}^{J/\psi\omega}}= \frac{1}{(2\pi)^3}\frac{1}{4M_{B_c}^2} k^{\prime}
\tilde{p}_{\omega}\Big( 3 |A|^2 |\vec{k}_{av}|^4 |t_1|^2 +\frac{2}{3} |B|^2 |\vec{k}|^4 |t_2|^2\Big)\, ,
\end{equation}
\begin{equation}
\frac{d\Gamma}{dM_{inv}^{D^*\bar{D}^*}}= \frac{1}{(2\pi)^3}\frac{1}{4M_{B_c}^2} k^{\prime}
\tilde{p}_{D^*}\Big( 3 |A|^2 |\vec{k}_{av}|^4 |t_3|^2 +\frac{2}{3} |B|^2 |\vec{k}|^4 |t_4|^2\Big)\, ,
\end{equation}
where $k^{\prime}$ is the $\pi$ momentum in the $B_c^-$ rest frame, $\tilde{p}_{\omega}$ the
$\omega$ momentum in the $J/\psi\omega$ rest frame and $k$ the pion momentum in the
$J/\psi\omega$ rest frame for the $J/\psi\omega$ final state,
\begin{equation}
k^{\prime} = \frac{\lambda^{1/2}(M^2_{B_c},m^2_{\pi},M_{inv}^{2\, J/\psi\omega })}{2M_{B_c}}\, ,
\end{equation}
\begin{equation}
k = \frac{\lambda^{1/2}(M^2_{B_c},m^2_{\pi},M_{inv}^{2\, J/\psi\omega })}{2M_{inv}^{J/\psi\omega}}\, ,
\end{equation}
\begin{equation}\label{ptil}
\tilde{p}_{\omega} = \frac{\lambda^{1/2}(M_{inv}^{2\,J/\psi\omega},
M^2_{J/\psi},m^2_{\omega})}{2 M_{inv}^{J/\psi\omega}}\, .
\end{equation}
For the $D^*\bar{D}^*$ final state in $k,\,k^{\prime}$ we change
$M_{inv}^{J/\psi\omega}$ to $M_{inv}^{D^*\bar{D}^*}$ and $\tilde{p}_{D^*}$ is like $\tilde{p}_{\omega}$
changing also $M_{inv}^{J/\psi\omega}$ by $M_{inv}^{D^*\bar{D}^*}$ and $M_{J/\psi}$, $m_{\omega}$ by
$M_{D^*}$, $M_{\bar{D}^*}$.

We get the amplitudes $t^I$ and $t^{II}$ from Ref.~\cite{molina}. We take them using the Flatt\'e form
of the amplitude in terms of the couplings obtained in Ref.~\cite{molina} and the width. The couplings
are given in Table \ref{coup0}.
\begin{table}[h]
 \begin{center}
 \caption{Couplings $g_{i}$ of the $0^{++}$ and $2^{++}$ resonances to the relevant channels, in units of MeV.}
\label{coup0}
\centerline{$\sqrt{s}_{pole}=3943 + i 7.4$, $I^G[J^{PC}]=0^+[0^{++}]$}
\vspace{0.2cm}
\begin{tabular}{ccccc}
\hline
~~~~~~ $D^*\bar{D}^*$~~~~~~ &~~~~~~ $D^*_s\bar{D}_s^*$~~~~~~ &~~~~~~   $K^*\bar{K}^*$ ~~~~~~  & ~~~~~~ $\rho\rho$ ~~~~~~ &~~~~~~  $\omega\omega$~~~~~~ \\
\hline
\hline
$18810-i 682 $ & $8426+i 1933 $ & $10- i 11$ & $-22 + i 47$ & $1348+ i 234 $\\
\hline
\end{tabular}\\
\vspace{0.2cm}
\begin{tabular}{ccccc}
\hline
~~~~~~ $\phi\phi$~~~~~~ &~~~~~~ $J/\psi J/\psi$~~~~~~ &~~~~~~ $\omega J/\psi$~~~~~~ &~~~~~~ $\phi J/\psi$~~~~~~ &~~~~~~ $\omega\phi$~~~~~~ \\
\hline
\hline
$-1000 -i 150$&$417+ i 64$&$-1429 - i 216$&$889+ i 196 $&$-215 - i107$\\
\hline
\end{tabular}
\end{center}
\begin{center}
\centerline{$\sqrt{s}_{pole}=3922+i 26$, $I^G[J^{PC}]=0^+[2^{++}]$}
\vspace{0.2cm}
\begin{tabular}{ccccc}
\hline
~~~~~~ $D^*\bar{D}^*$~~~~~~ &~~~~~~ $D^*_s\bar{D}_s^*$~~~~~~ &~~~~~~ $K^*\bar{K}^*$~~~~~~ &~~~~~~ $\rho\rho$~~~~~~ &~~~~~~ $\omega\omega$~~~~~~ \\
\hline
\hline
$21100- i 1802 $&$1633+i 6797 $&$42+ i 14 $&$-75 +i 37$&$1558 + i 1821$\\
\hline
\end{tabular}\\
\vspace{0.2cm}
\begin{tabular}{ccccc}
\hline
~~~~~~ $\phi\phi$~~~~~~ &~~~~~~ $J/\psi J/\psi$~~~~~~ &~~~~~~ $\omega J/\psi$~~~~~~ &~~~~~~ $\phi J/\psi$~~~~~~ &~~~~~~ $\omega\phi$~~~~~~ \\
\hline
\hline
$-904 - i1783 $&$1783 +i 197$&$-2558 - i2289$&$918+ i2921 $&$91 -i 784$\\
\hline
\end{tabular}
\end{center}
\end{table}

The amplitudes are given by
\begin{equation}
t^{i}_{D^*\bar{D}^*,\, j} = \frac{g^{(i)}_{R,\,D^*\bar{D}^*}\,\, g^{(i)}_{R,\,j}}{M^{2\,\,j}_{inv}
 - M^2_{R_i} + i M_{R_i}\Gamma_{R_i}} \, ,
\end{equation}
with $i=I,\,II$, and $j=J/\psi\omega$ or $D^*\bar{D}^*$. We also have
\begin{equation}
t^{i}_{D_s^*\bar{D}_s^*,\, j} = \frac{g^{(i)}_{R,\,D_s^*\bar{D}_s^*}\,\, g^{(i)}_{R,\,j}}{M^{2\,\,j}_{inv}
 - M^2_{R_i} + i M_{R_i}\Gamma_{R_i}} \, ,
\end{equation}
where the width is taken as
\begin{equation}\label{gamma}
\Gamma_{R_i} = \Gamma^{(i)}_0 + \Gamma^{(i)}_{J/\psi\omega} + \Gamma^{(i)}_{D^*\bar{D}^*}\, ,
\end{equation}
with
\begin{equation}
\Gamma_{J/\psi\omega}^{(i)} = \frac{|g^{i}_{R,\,J/\psi\omega}|^2}{8\pi M^2_{R_i}}\,\tilde{p}_{\omega}\, ,
\end{equation}
and $\tilde{p}_{\omega}$ given by Eq.~\eqref{ptil} as a function of $M_{inv}^{J/\psi\omega}$
or $M_{inv}^{D^*\bar{D}^*}$ depending on the reaction studied, and
\begin{equation}\label{gammadd}
\Gamma_{D^*\bar{D}^*}^{(i)} = \frac{|g^{i}_{R,\,D^*\bar{D}^*}|^2}{8\pi M^2_{R_i}}\,\tilde{p}_{D^*}
\Theta(M_{inv} - 2 M_{D^*})\, ,
\end{equation}
with $\tilde{p}_{D^*}$ as $\tilde{p}_{\omega}$ in Eq.~\eqref{ptil} with the changes $M_{J/\psi}\to M_{D^*}$,
$M_{\omega}\to M_{\bar{D}^*}$, and $M_{inv}\to M_{inv}^{J/\psi\omega}$ or $M_{inv}^{D^*\bar{D}^*}$
depending on the reaction studied. The width $\Gamma^{(i)}_0$ in Eq.~\eqref{gamma} accounts for the channels
different of $J/\psi\omega$ and $D^*\bar{D}^*$, mostly the light channels, such that $\Gamma^{(i)}_0$
is practically constant and we take
\begin{equation}
\Gamma^{(i)}_0 = \Gamma_{R_i} - \Gamma^{(i)}_{J/\psi\omega}(M_{inv}^{J/\psi\omega}=M_{R_i})\, .
\end{equation}

Note that in Eq.~\eqref{gammadd}, $\Gamma^{(i)}_{D^*\bar{D}^*}$ only starts above the
$D^*\bar{D}^*$ threshold, but since the coupling of the resonance to this channel is so large, it
grows fast above threshold giving rise to the Flatt\'e effect.

\section{Results}

We will present the invariant mass distribution in arbitrary units, but $\frac{d\Gamma}{dM_{inv}^{J/\psi\omega}}$
and $\frac{d\Gamma}{dM_{inv}^{D^*\bar{D}^*}}$ will have the same normalization. For this purpose, we take
$A=1$ and look at the results for different values of $B$. Since $A$ and $B$ have been normalized to have
the same dimensions, providing similar strength for the two terms for $A=B$, we will take values of $B$
close to $1$, $0.5$, $1$, $1.5$ and $2$. We show in Fig. 5 the results of $\frac{d\Gamma}{dM_{inv}^{J/\psi\omega}}$
and in Fig. 6 for $\frac{d\Gamma}{dM_{inv}^{D^*\bar{D}^*}}$ for these different values. The absolute normalization
is arbitrary and the shape changes a bit since one give more strength to one or another resonance changing $B$.

\begin{figure}
	\begin{center}
		\includegraphics[width=0.65\textwidth]{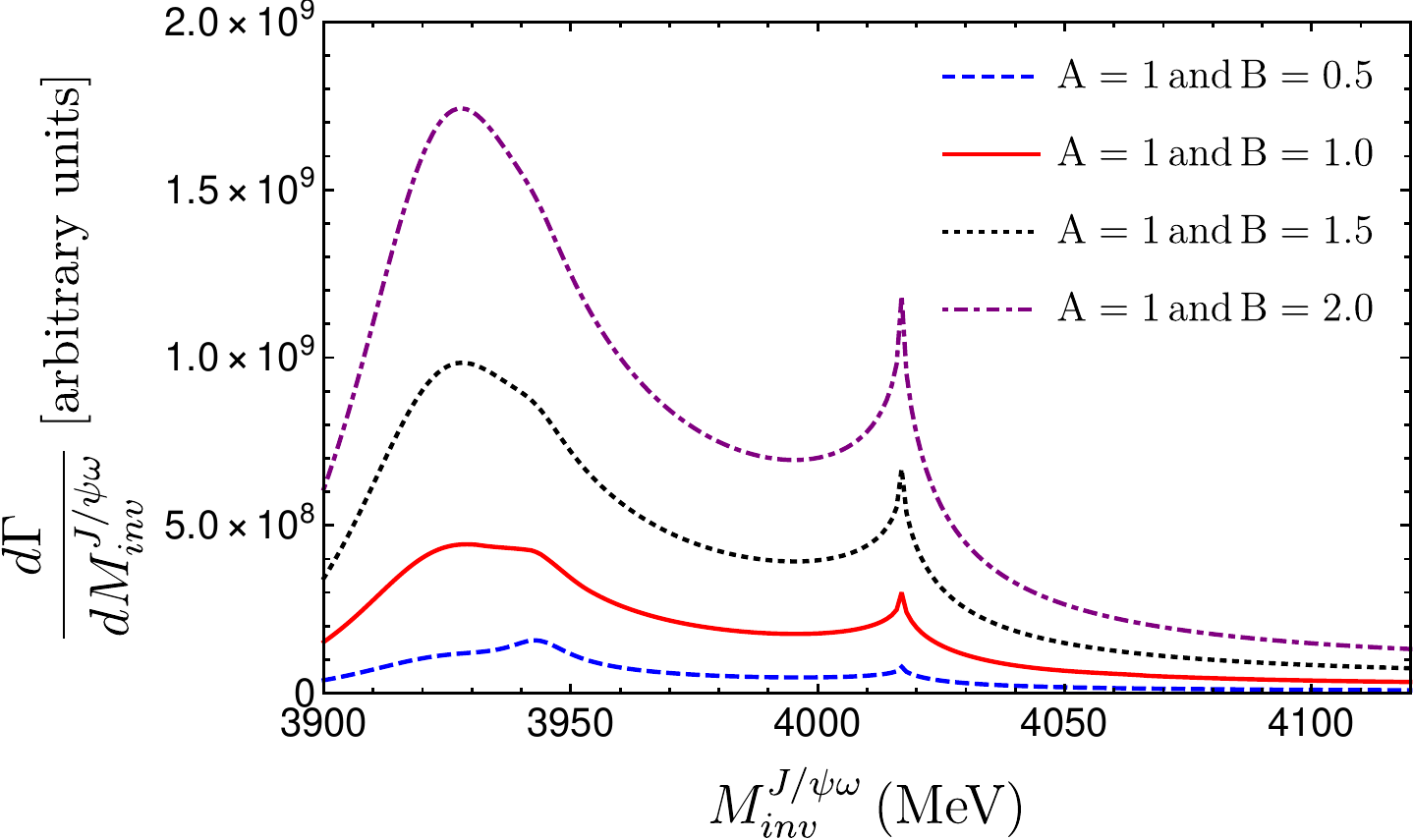}
	\end{center}
	\caption{\label{dgammaj} $\frac{d\Gamma}{dM_{inv}^{J/\psi\omega}}$ the results for the different values of the
	parameter $B$.}
\end{figure}

\begin{figure}
	\begin{center}
		\includegraphics[width=0.65\textwidth]{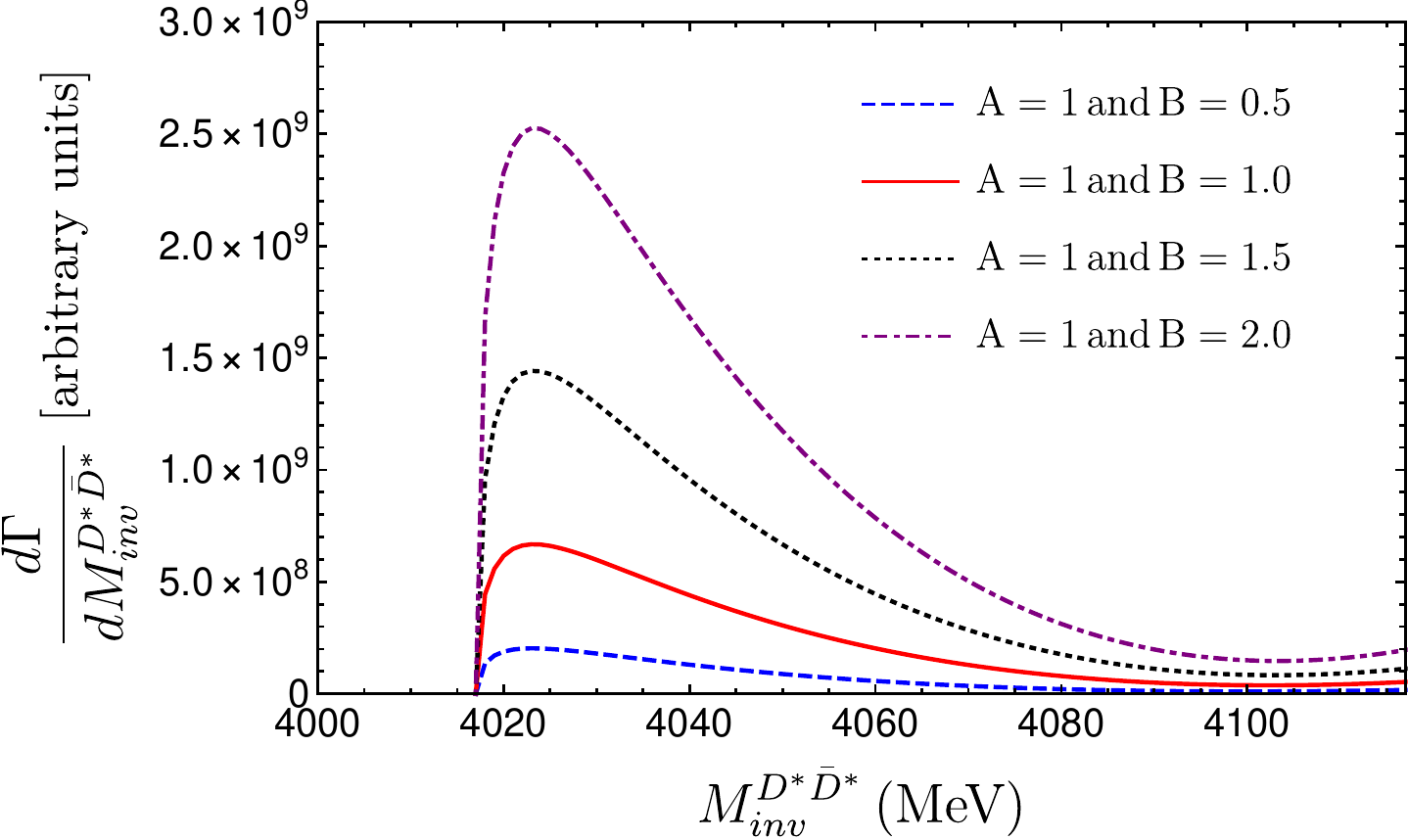}
	\end{center}
	\caption{\label{dgammaDD} $\frac{d\Gamma}{dM_{inv}^{D^*\bar{D}^*}}$ the results for the different values of the
	parameter $B$. The normalization is the same as in Fig.~\ref{dgammaj}.}
\end{figure}

In Fig.~\ref{dgammaj} we see that due to the proximity of the two resonances, and
the fact that both of them can be produced in this reaction, the two peaks actually
merge into a broader one, although a precise measurement could maybe allow a
separation of the two peaks, particularly if a partial wave analysis is done that separates
the two different spin resonances. Interesting, however, is the fact that the cusp appears
always at the same place, the $D^*\bar{D}^*$ threshold. The other relevant feature is that
its strength grows with increasing weight of the tensor resonance, indicating that the cusp
is basically tied to the $2^{++}$ $X(3930)$ state.

In Fig.~\ref{dgammaDD} we show the $D^*\bar{D}^*$ mass distribution in the
$B_c^-\to \pi^- D^*\bar{D}^*$ decay. We observe a distribution quite different
from ordinary phase space, sticking close to threshold, indicating that it is
influenced by a resonance below threshold. Its strength also grows with increasing
strength of the tensor resonance, which establishes a link between this state
and the $D^*\bar{D}^*$ distribution.

Very interesting is the ratio of the strengths of $\frac{d\Gamma}{dM_{inv}^{J/\psi\omega}}$ at the
peak of the cusp of the $D^*\bar{D}^*$ threshold versus the strength at the peak of
$\frac{d\Gamma}{dM_{inv}^{D^*\bar{D}^*}}$. We show these numbers in Table~\ref{tab2}
for different values of $B$. As we can see, this ratio is relatively stable and tied to the
dynamically generated nature of the two resonances discussed.

\begin{table}[h!]
\caption{Ratio $R$ between the maximum of the $D^*\bar{D}^*$ mass distribution in
Fig.~\ref{dgammaDD} and the strength of the cusp at the $D^*\bar{D}^*$ threshold in
Fig.~\ref{dgammaj}.}
\centering
\begin{tabular}{c | c | c | c  }
\hline\hline
$A=1.0$ and $B=0.5$ ~& ~$A=1.0$ and $B=1.0$ & ~$A=1.0$ and $B=1.5$~ & ~$A=1.0$ and $B=2.0$\\
\hline
$R=2.57$ ~& $R=2.22$ & $R=2.16$ & $R=2.13$\\
\hline\hline
\end{tabular}
\label{tab2}
\end{table}

The fact that the ratio $R$ is essentially independent on the strength $B$ of the
tensor resonance indicates that it is this resonance in practice the one that is
responsible for both the cusp in the $J/\psi\omega$ and the $D^*\bar{D}^*$ mass
distributions in the $B_c^-\to \pi^- D^*\bar{D}^*$ reaction.

\section{Conclusions}
\label{sec:conc}

We have looked into the $B_c^-\to J/\psi \omega$ decay and in particular in the $J/\psi \omega$
mass distribution. We find that this observable is much influenced by the role of the
$X(3940)$ and $X(3930)$ resonances, which in Ref.~\cite{molina} appear dynamically generated
from the vector-vector meson interaction in the charm sector. These resonances couple mostly to
$D^*\bar{D}^*$ in $2^{++}$ and $0^{++}$, respectively. In order to find support for this nature
of the resonances we stress two particular features: the first one is to observe that $J/\psi\omega$ is
not the main channel for this resonances, but $D^*\bar{D}^*$. As a consequence, one finds a strong
cusp at the $D^*\bar{D}^*$ threshold in the $J/\psi\omega$ mass distribution. The other feature is that
since the resonances are tied to $D^*\bar{D}^*$, they should influence the $D^*\bar{D}^*$ mass
distribution close to threshold in the $B_c^-\to \pi^- D^*\bar{D}^*$ reaction. What we find is that, within
uncertainties tied to our ignorance of the weight by which the $X(3940)$ and $X(3930)$ resonances
are produced, the ratio of the strength at the cusp peak and the strength at the maximum of the
$D^*\bar{D}^*$ mass distribution are related and quite independent of the relative weight of these
two resonances. This is because the $D^*\bar{D}^*$ mass distribution is more influenced by the
$X(3930)$ resonance that has a larger width.

In addition we observe also a peak around $3930-3940$ MeV in the $J/\psi\omega$ mass
distribution, corresponding to the excitation of these two resonances, and show that the
cusp at the $D^*\bar{D}^*$ threshold has similar strength as the peak. All these features,
when observed, should serve to support the molecular nature of these resonances and we can only
encourage the performance of the experiments.

\section*{Acknowledgements}

L. R. Dai wishes to acknowledge the support from the
State Scholarship Fund of China (No. 201708210057) and
the National Natural Science Foundation of China
(No. 11575076). J. M. Dias thanks the Funda\c c\~ao de
Amparo \`a Pesquisa do Estado de S\~ao Paulo (FAPESP)
for support by FAPESP grant 2016/22561-2. This work is partly
supported by the Spanish Ministerio de Economia y Competitividad
and European FEDER funds under the contract number FIS2011-
28853-C02-01, FIS2011- 28853-C02-02, FIS2014-57026-
REDT, FIS2014-51948-C2- 1-P, and FIS2014-51948-C2-
2-P, and the Generalitat Valenciana in the program
Prometeo II-2014/068 (EO).

\end{document}